# Natural User Interfaces: Trend in Virtual Interaction

Dr. Manju Kaushik[1] Rashmi Jain[2]

[1]*Associate Professor, Computer Science and Engineering, JECRC University*

[2]*Research Scholar, Computer Science and Engineering, JECRC University*

[1]`Email:manju.kaushik@jecrcu.edu.in`

[2]`Email:jainrashmi17@gmail.com`

*Abstract*

**Based on the fundamental constraints of natural way of interacting such as speech, touch, contextual and environmental awareness, immersive 3D experiences-all with a goal of a computer that can see listen, learn talk and act. We drive a set of trends prevailing for the next generation of user interface: Natural User Interface (NUI).New technologies are pushing the boundaries of what is possible without touching or clicking an interface- paving the way of interaction to information visualization and opportunities in human towards more natural interaction than ever before. In this paper we consider the trends in computer interaction through that must be taken into consideration to come up-in the near future with a well-designed-NUI.**

## I. INTRODUCTION

Rapid improvements in computer technology have increased the opportunities for developing highly interactive user interfaces .Natural User Interface (NUI), a more natural way for people to interact with technology. NUI refers to both sensory inputs such as touch, speech and gesture but goes much further to describe computing that is intelligent and contextually aware, with the ability to recognize a person's face, environment and intent, even emotions and relationships. Computers are now taken on roles in almost all aspects of life. People and lifestyles are changing. These changes are sometimes spurred on by technology, but other times work in parallel or provoke technological innovation. There is a global scale of change which is taking place hand in hand with new technologies. In this Paper, we provide trends and study to pave the path towards future. There have been various computer-driven revolutions in the past: started with the introduction of the personal computer (PC), then invention of the graphical browser, and the giant network of networks Internet. A true revolution in data-input methods will come when we move from GUI (graphic user interfaces) to NUI (natural user interfaces), from mouse and keyboard to speech and gesture. A system with a NUI supports the mix of real and virtual objects. As input it recognizes (visually, acoustically or with other sensors) and understands physical objects and humans acting in a natural way (e.g., speech input, handwriting, etc.). Its output is based on pattern projection such as video projection, holography, speech synthesis or 3D audio patterns. A necessary condition in our definition of a NUI is that it allows inter-referential I/O i.e. that the same modality is used for input and output interface requirements. Natural User Interface (NUI) has three major trends: multi-touch, voice, and gesture interaction. These NUI trends manifest in multiple form factors and developer technologies and all can leverage innovative capabilities like multi-user, cloud, and parallel computing [1].

## II.TRENDS IN NUI

The interaction between digital technologies and the physical objects that are embedded in will change existing forms of interaction. As Human Beings, yearn for a more interactive, intuitive, and lively method of communicating with the "digital-world." We want to be able to interact with data objects in the same way we interact with physical objects--even for those who fear losing conventional methods of communication with technology such as "mouse and keyboard," natural interfaces are attractive in that their emulation of everyday "real-world" gestures perfectly match our envisionment for how we believe technology should work. For example, with Multi-touch interfaces, we are allowed to treat our data collections as a "workspace." In this sense a computer evolves beyond what computers were known to be and become analogous to physical tools. Major trends in interaction prevailing had lessen the gap between real and virtual objects. We will need new conceptual models and metaphors of how best to support and control these new forms of more 'natural' but less obvious forms of interaction [2]. Research is needed to determine what will be the most natural, efficient and socially accepted means of controlling such interactions.

A. *Multi Touch Interaction*

Multi-touch technology can be simply divided into two parts: hardware and software. Hardware serves to complete the information collection and software to complete the analysis of information which are finally converted into specific user command [3]. It is believed that the Multi-touch key technology should include the following major components: Multi-touch Hardware Platform: These platforms have their own advantages and disadvantages. Study of these platforms helps to understand how to build interaction platforms of lower cost, more convenient installation and more precise target selection and to study a number of other interactive technology unrelated to the platforms. The Accuracy of Selection for Multi-touch Device Precision choice technology, in fact is the detection of contact tracing, and it has great significance on how to accurately track and locate contacts to achieve the freedom of gesture interaction. In particular, when the target size is very small, how our fingers could accurately locate the goal we want, is the content worth deep study.

B. *Voice Interaction*

*No longer has the sovereign property of humans, speech become an ability we share with machines.*
— Sarah Borruso

Voice interfaces have a range of labels — as above, many of them are configurations of "spoken," "dialogue," and "system," with each other or related words, usually reduced to acronyms (see the glossary for the most common terms). *VUI* (Voice User Interface) has recently emerged as the leading short-hand term. They work on a linguistic paradigm (word strings), and consist of utterances, plain and simple: speech in and speech out. In attempting to develop a conversational interface there are a number of interface requirements [4] these can be divided into two groups; requirements on the interface of the conversational system, and requirements on the underlying architecture. Considering human face to face conversation it is obvious that a conversational system must be able to recognize and respond to verbal and non-verbal input and be able to generate verbal and non-verbal output. This necessitates some form of graphical representation, both to display non-verbal cues as well as to act as a point of reference for the many spatial cues that occur in conversation. Without an Embodied interface it is impossible to look your computer in the eye to tell it to take the turn in the conversation!

C. *Gesture Interaction*

If we remove ourselves from the world of computers and consider human-human interaction for a moment we quickly realize that we utilize a broad range of gesture in communication. The gestures that are used vary greatly among contexts and cultures yet are intimately related to communication.

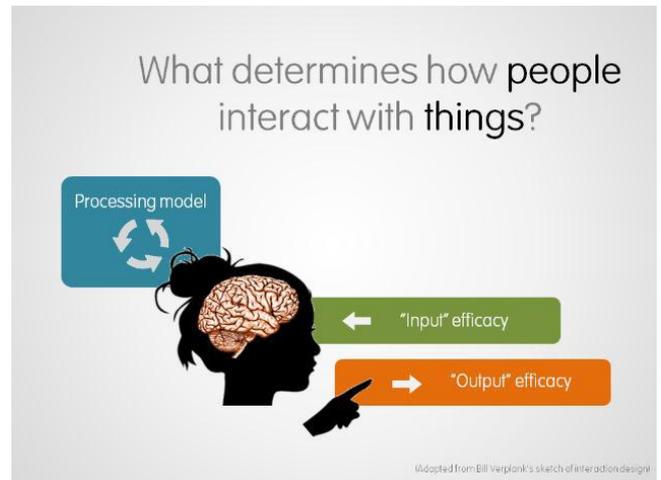

Bill Verplank's model of HCI describes user interactions with any system as a function of three human factors [5]
1. *Input efficacy*: how and how well we sense what a system communicates to us?
2. *Processing model*: how and how well we understand and think about that communication and system's functionality and behavior?
3. *Output efficacy:* how and how well we communicate back to the system?

This is shown by the fact that people gesticulate just as much when talking on the phone and can't see each other as in face to face conversation Gestures can exist in isolation or involve external objects. Free of any object, we wave, beckon, fend off, and to a greater or lesser degree (depending on training) make use of more formal sign languages. With respect to objects, we have a broad range of gestures that are almost universal, including pointing at objects, touching or moving objects, changing object shape, activating objects such as controls, or handing objects to others. This suggests that gestures can be classified according to their function. Uses function to group gestures into three types [7]:

- Semiotic: those used to communicate meaningful information.
- Ergotic: those used to manipulate the physical world and create artifacts
- Epistemic: those used to learn from the environment through tactile or haptic exploration

The gestural equivalents of direct manipulation interfaces are those which use gesture alone. These can range from interfaces that recognize a few symbolic gestures to those that implement fully fledged sign language interpretation [8-11]. Similarly interfaces may recognize static hand poses, or dynamic hand motion, or a combination of both. In all cases each gesture has an unambiguous semantic meaning associated with it that can be used in the interface. The advantages of the gesture- based interaction design have been highlighted as the following:

- It provides a simple, usable and interesting user interface and satisfies the need for more freedom in a human computer interaction environment.
- The expectations of the users, the cognitive and psychological design aspects of the gesture- based interaction technology are met perfectly and an easy to understand and use.
- It provides people new experience and great pleasure which traditional interaction could not offer.
- It makes the interaction between human and computer more natural. It has been illustrated in science fiction movies that this technology can improve people's lives if it is applied rightly.
- It is considered as a powerful tool for computers to begin to understand human body language thus building a richer bridge between machines and humans than primitive text user interfaces or graphical user interfaces (GUI), which still limit the majority of input to keyboard and mouse.
- It is widely used in various application areas since it gives the user a new experience of feeling.

Related approaches that support gesture-based interaction have been developed in various application areas, such as sign language, navigation system, medical research, robot control, browsing, game applications, and augmented reality applications. User-centered design is a cornerstone of human-computer interaction. But users are not designers; therefore, care must be taken to elicit user behavior profitable for design.

## III. CONCLUSION

In this paper we have study the trends in Natural User Interface with the computer systems becoming more sophisticated, More are beginning to make choices and decisions on our behalf; computers become more autonomous they also have become increasingly present in our world. More research can follow the upcoming trends.